\documentclass[]{emulateapj}

\usepackage{natbib}
\usepackage{color}
\newcommand{\eqref}[1]{(\ref{#1})}

\shorttitle{}
\shortauthors{}

\begin{document}

\title{The Impact of Cluster Structure and Dynamical State on Scatter in the Sunyaev-Zel'dovich Flux-Mass Relation}

\author{H.-Y.\ Karen Yang\altaffilmark{1},
Suman Bhattacharya\altaffilmark{2},
and Paul M. Ricker\altaffilmark{1,3}}
\altaffiltext{1}{Department of Astronomy, University of Illinois, Urbana, IL 61801}
\altaffiltext{2}{Los Alamos National Laboratory, Los Alamos, NM}
\altaffiltext{3}{National Center for Supercomputing Applications, Urbana, IL 61801}
\email{hyang20@illinois.edu, pmricker@illinois.edu, sumanb@lanl.gov}

\begin{abstract}

Cosmological constraints from cluster surveys rely on accurate mass estimates from the mass-observable relations. In order to avoid systematic biases and reduce uncertainties, we study the form and physical origin of the intrinsic scatter about the mean Sunyaev-Zel'dovich (SZ) flux-mass relation using a hydrodynamical simulation of galaxy cluster formation. We examine the assumption of lognormal scatter and detect non-negligible positive skewness and kurtosis ($>0.5$) for a wide range of limiting masses and redshifts. These higher-order moments should be included in the parametrization of scatter in order not to bias cosmological constraints. We investigate the sources of the scatter by correlating it with measures of cluster morphology, halo concentration, and dynamical state, and we quantify the individual contribution from each source. We find that statistically the impact of dynamical state is weak, so the selection bias due to mergers is negligible. On the other hand, there is a strong correlation between the scatter and halo concentration, which can be used to reduce the scatter significantly (from $12.07\%$ to $7.34\%$ or by $\sim 40\%$ for clusters at $z=0$). We also show that a cross-calibration by combining information from X-ray followups can be used to reduce the scatter in the flux-mass relation and also identify outliers in both X-ray and SZ cluster surveys.
\end{abstract}



\section{Introduction}

The evolution of structure in the Universe is thought to be a
hierarchical process driven by gravitational instability acting on
primordial density fluctuations. In this dynamical process, smaller
clumps of matter merge to form bigger ones within a ``cosmic web'' of
spatial structure incorporating matter within tenuous sheets, higher-density
filaments, and large matter concentrations at the nodal points
of the web. Clusters of galaxies occupy the top of the mass
hierarchy, being the largest objects that have had time to collapse and
form due to their self-gravity. Therefore, they are the most recent structures
to form. Clusters have two very important features:
they are very massive ($10^{14-15} M_\odot$) and populate the exponential tail 
of the mass function, which is sensitive to dark energy \citep{haiman01}, and
because of their deep potential wells, they are ``matter traps,''
preventing their internal constituents from escaping. The first aspect,
combined with their young age, makes clusters an excellent probe of
cosmology, being especially sensitive to dark matter and dark
energy. The second makes clusters excellent laboratories for
studying key astrophysics problems such as star and galaxy
formation. Because of their central importance to both cosmology and
astrophysics, clusters are being targeted by a variety of
observational programs.

Clusters can be detected optically via starlight from their member galaxies, in
X-ray due to thermal emission from hot gas, and using the thermal Sunyaev-Zel'dovich (SZ) effect \citep{1972CoASP...4..173S}. The SZ effect arises from the inverse-Compton upscattering of cosmic microwave background (CMB) photons by the hot electrons in a cluster (several keV), leading to an increase of the CMB brightness
temperature at frequencies above about 250~GHz and a decrease at lower frequencies. This distortion of the Planck spectrum can be measured by
sensitive, high-resolution CMB telescopes from the ground \citep{act, spt09, spt10}, and it is 
one of the best ways to find clusters at higher redshifts \citep{spt-hiz}, since the SZ effect is (almost) redshift-independent.

Upcoming surveys will detect clusters using their SZ signatures. However, instead of mass $M$, SZ surveys measure $Y$, the SZ Compton optical depth (the so-called $y$-decrement) integrated over a portion of a cluster's projected area. To obtain the cluster mass distribution from surveys, we need a so-called mass-observable ($Y$--$M$) relation.
Using the mass distribution to constrain cosmological parameters requires that
we know the errors introduced in this process, so we must also quantify the scatter and the bias in the mass-observable relation as functions of redshift \citep{lima04, lima05, 2010PhRvD..81h3509C}.

Because of the exponential shape of the cluster mass function, scatter in the $X$--$M$ relation (for any observable $X$ that is positively correlated with $M$) boosts the number density
of clusters observed in logarithmic bins of $X$, as the overall number of lower-mass
clusters scattering to higher values of $X$ far exceeds the number of high-mass clusters scattering
in the opposite direction.  As shown in \cite{bh10}, a few percent systematic error in mass, which can arise due to bias in the $X$--$M$ relation, leads to a significant difference in the tail of the mass function. Thus, misestimating the scatter and bias in the $X$--$M$ relation can lead to biases in the cosmological constraints \citep[e.g.][]{2002ApJ...577..579R, 2008ApJ...680...17W}. Moreover, previous work forecasting cosmological constraints based on cluster surveys has assumed the distribution of scatter to be lognormal  \citep{lima05, 2009PhRvD..79f3009C}. If this assumption were invalid, it would also lead to biases in the results \citep{shaw10}.

Understanding the physical sources of scatter can help us to reduce this scatter and improve the mass estimates. For example, the use of core-excised quantities reduces the effects of cool cores in clusters and hence also the large scatter in the X-ray luminosity-temperature ($L_X$--$T_X$) relation \citep{1998MNRAS.297L..57A} or the $L_X$--$M$ relation \citep{2010MNRAS.406.1773M}.
Using the halo concentration as a third parameter can reduce the scatter in the $T_X$--$M$ relation \citep{2009ApJ...699..315Y}. Combinations of observables with oppositely trending scatter can also help, as shown by the tight correlation between the X-ray
counterpart of the Compton $y$-parameter, $Y_X \equiv M_{\rm gas}T_X$, and mass \citep{2006ApJ...650..128K}. These examples illustrate the possibility of obtaining better mass estimates if our knowledge of the physical origin of scatter can be improved.

Because of spatial resolution constraints, the $Y$--$M$ relation has been studied using two types of hydrodynamic simulations. One type includes only adiabatic physics but has sufficient statistics to quantify the scatter and the bias in the mass-observable relation \citep{santafe06}. The other includes extra physics such as radiative cooling, star formation, and supernova feedback \citep{nagai07} or feedback from quasars \citep{quasar07}, but with limited statistics (16 and 10 halos, respectively, in the cited references). Another approach is to include gas physics using a semianalytic gas prescription in halos obtained from a dark-matter only (DMO) simulation \citep[e.g.][]{bode07, bode09, shaw10a}. In particular, \cite{shaw07, shaw10} have quantified the $Y$--$M$ relation and its non-gaussianity using the ``DMO+semianalytic'' approach. As pointed out by \cite{shaw07}, the hydrodynamic simulations tend to show slightly different scatter compared to the ``DMO+semianalytic'' case especially for overdensities $\sim$500 times the critical density of the universe. These differences need to be understood.  Only recently have there been attempts to incorporate extra baryonic physics into a large cosmological simulation to study the intrinsic variances in the scaling relations. For instance, \cite{2010ApJ...715.1508S} have used resimulations from the Millennium Gas Simulation with gas dynamics treated in both gravity-only and cooling plus preheating prescriptions to study the scaling relations and correlations among cluster structural properties and observables in X-ray and SZ. 

In this study we begin a systematic investigation of the physical sources of the intrinsic $Y$--$M$ scatter and their impact on the form of scatter for the purpose of improving our knowledge of both cluster formation and cluster cosmology. We use cosmological simulations to obtain clusters with sufficient statistics and to incorporate hydrodynamic processes including interactions with large-scale environment, variations in cluster structures, and merger-induced shock heating and departure from hydrostatic equilibrium. Since we would like to focus on the scatter driven by gravitational effects only, radiative cooling and heating mechanisms are not included. We will address the influence of radiative cooling and feedback explicitly in a separate paper. In this work, we examine the assumption of lognormal scatter, which has important implications for self-calibration studies of cluster surveys. We investigate various sources of scatter, such as halo concentrations, dynamical state,
and cluster morphology, by correlating the scatter with quantitative measures of each source. We show that these correlations can be used to reduce the scatter and tighten the scaling relation. We also discuss possible applications and issues when combining SZ and X-ray scaling relations.

The outline of this paper is as follows. In \S~\ref{Sec:method} we summarize the key components of our simulation and analyses, including the numerical methods and simulation parameters, a brief overview of the SZ effect, merger tree construction, and how we create idealized cluster samples to disentangle the sources of scatter. The $Y$--$M$ relation and the form of its scatter are presented in \S~\ref{Sec:result}. Possible sources of scatter are investigated in \S~\ref{Sec:source}. In \S~\ref{Sec:xray} we explore the possibility of combining SZ and X-ray scaling relations to improve cluster mass estimates. Finally, we discuss our results and give the conclusions in \S~\ref{Sec:conclusion}.


\section{Method}
\label{Sec:method}

\subsection{Simulation}

The simulation described here was performed using FLASH, an Eulerian hydrodynamics plus $N$-body code which has been applied to a wide range of problems and extensively validated for hydrodynamical \citep{2002ApJS..143..201C} and cosmological $N$-body \citep{2005ApJS..160...28H, 2008CS&D....1a5003H} applications. We used version 2.4 of FLASH together with the local transform-based multigrid Poisson solver described by \cite{2008ApJS..176..293R}. Because we are concerned in this paper only with the effect of gravity-driven processes on mass-observable relations, the calculation described here
did not employ radiative cooling or feedback due to star formation or active galaxies. Here we give a brief summary and refer the readers to \cite{2009ApJ...699..315Y} for details of the numerical methods and merger tree analysis. 

The results presented here are based on a FLASH simulation of structure formation in
the $\Lambda$CDM cosmology within a 3D cubical volume spanning $256 h^{-1}$~Mpc.
Initial conditions were generated for a starting redshift $z = 66$ using GRAFIC
\citep{grafic} with an initial power spectrum generated using CMBFAST \citep{cmbfast}.
The cosmological parameter values used were chosen to be consistent with the third-year
WMAP results \citep{wmap3}: present-day matter density parameter $\Omega_{m0} = 0.262$,
present-day baryonic density parameter $\Omega_{b0} = 0.0437$, present-day cosmological
constant density parameter $\Omega_{\Lambda 0} = 0.738$, matter power spectrum
normalization $\sigma_8 = 0.74$, and Hubble constant $h = 0.708$ ($H_0 = 100h$~km~s$^{-1}$~Mpc$^{-1}$). The simulation contains $1024^3$ dark matter particles
with a particle mass $m_p = 9.2\times10^8 h^{-1} M_\odot$. The mesh used for the gasdynamics
and potential solution was fully refined to $1024^3$ zones, which corresponds to a comoving
zone spacing of $250 h^{-1}$~kpc. Considering the effect of resolution on the computed
abundances of halos of different mass \citep{2007ApJ...671.1160L},
with these parameters we are able to capture all
halos containing more than 3150 particles (i.e.\ total mass $2.9\times10^{12} h^{-1} M_\odot$)
and 1150 particles (i.e.\ $1.1\times10^{12} h^{-1} M_\odot$) at $z = 0$ and $z = 1$,
respectively. The halos are identified using the friends-of-friends (FOF) algorithm. The overdensity
mass and radius, $M_\Delta$ and $R_\Delta$, are then found by growing spheres around each FOF center until the averaged total density is $\Delta$ times the {\it critical} density of the universe. The simulation was carried out using 800 processors of the Cray XT4 system at Oak Ridge
National Laboratory, requiring a total of 16,500 CPU-hours. 


\subsection{Sunyaev-Zel'dovich Effect}

The thermal Sunyaev-Zel'dovich effect is caused by inverse Compton scattering of CMB photons off the hot electrons inside galaxy clusters. Assuming that relativistic corrections are small, the resulting distortion of the CMB temperature can be written as $\Delta T/T_{CMB}=g_\nu(x)y$ where $g_\nu(x)=x(\coth(x/2)-4)$ with $x=h\nu/k_BT_{CMB}$, $k_B$ is the Boltzmann constant, $\nu$ is the frequency of observation, and $T_{CMB}$ is the mean CMB temperature at the current epoch. Here $y$ is the Compton $y$-parameter, which can be written as
\begin{equation}
y= \frac{k_B\sigma_T}{m_ec^2}\int n_e(l)T_e(l)dl,
\label{eq:ylos}
\end{equation}
where $n_e(l)$ is the electron density profile, $\sigma_T$ is the Thomson scattering cross section, and the integration is along the line of sight, which is defined to be the $x$-direction in the simulation box. 

Note, however, that the observable is the integral of the temperature distortion over the cluster's projection onto the sky. For a cluster at redshift $z$, it is given by
\begin{equation}
Y(M,z) = \frac{1}{d_A(z)^2} \frac{k_B\sigma_T}{m_ec^2}\int n_e(l)T_e(l)dV,
\label{eq:yint}
\end{equation}
where $d_A(z)$ is the angular diameter distance to the cluster, and the integration is over the volume of the cluster. In the literature, sometimes the factor $1/d_A^2$ is omitted from Eq.~\ref{eq:yint}. In this study, when we omit the factor $1/d_A^2$, we denote the temperature distortion as $Y(M)$; otherwise it is denoted as $Y(M,z)$. Note that $Y(M,z)$ is dimensionless, while the units of $Y(M)$ are $\mbox{Mpc}^2$. In the following sections in which we investigate the distribution and origin of intrinsic scatter, we adopt $Y(M)$ unless explicitly stated otherwise.

Following the convention in the literature, we denote $Y_\Delta$ as the SZ flux integrated out to certain overdensity radius $R_\Delta$. In the simulation box, cell-averaged information about the gas density and temperature is stored for each grid cell. Thus for each cluster the integrated SZ distortion is calculated using
\begin{equation}
Y_\Delta (M,z)=  \frac{1}{d_A(z)^2} \frac{k_B\sigma_T}{m_ec^2}\sum_{i, r_p\leq R_\Delta} n_{e,i}T_{e,i}\Delta V_{i},
\label{eq:ygrid}
\end{equation}
where the summation is over all the grid cells across the cluster volume within a cylinder of projected radius $R_\Delta$, and $\Delta V_{i}$ is the volume of each grid cell. 


\subsection{Merger Tree Analysis}

In order to explore the influence of cluster formation and merger history on the intrinsic scatter, we construct a merger tree for each cluster in our simulation in the following way. Our simulation generates snapshots that contain particle tags and positions every $100 h^{-1}$~Myr beginning at $z=2$. For each snapshot, all the groups with more than 10 particles are found using the FOF halo finder with linking length parameter $b = 0.2$. Between successive outputs at times $t = t_n$, we 
find the progenitors at time $t_{n-1}$ for all the halos at $t_n$ by tracing the particle tags, 
which are uniquely assigned to each particle at the beginning of the simulation. For each halo at $t_n$ we record the masses of its progenitors, the masses they contributed to the halo, and the number of unbound particles. Then the merger trees are constructed by linking all the progenitors identified in the previous outputs for
halos above our halo completeness limit at $z=0$. Deriving the mass accretion histories
is straightforwardly accomplished by following the mass of the most massive progenitor back in time. Cluster formation time is often defined as the epoch when a cluster exceeds a certain fraction of its final mass. The commonly-adopted thresholds include 10\%, 25\%, 50\%, and 70\%. In discussion below we present results using the 50\% threshold.

To directly quantify the dynamical state of clusters without relying on morphology, we find the time since last merger for each cluster in our simulation. In our analysis, mergers are defined in two ways: the mass-jump definition, in which a merger is present if there is a mass jump larger than some threshold in the halo's assembly history; and the mass-ratio definition, which identifies a merger if the ratio of contributed masses from the first- and second-ranked progenitors is less than a certain value \citep{2005APh....24..316C}. To study the variations in cluster observables induced by different types of mergers, we use $1.2$ and $1.33$ as thresholds for the mass-jump definition and 10:1, 5:1, and 3:1 in the mass-ratio definition. In this paper, by `merging clusters' at a given lookback time we will refer to those identified by at least one of these five merger diagnostics in the preceding 3 Gyr, 
chosen to be long enough such that mergers with different impact parameters and mass ratios would have returned to virial equilibrium within $R_{500}$ \citep{2006MNRAS.373..881P}.
The mergers are `major' if the mass jump is larger than 1.2 or if the mass ratio is less than 5:1; `minor' mergers, on the other hand, have mass ratios between 10:1 and 5:1.


\begin{table*}[thdp]
\caption{Summary of models used for constructing the idealized cluster samples.}
\begin{center}
\begin{tabular}{ccc}
\hline
\hline
Sample & Assumptions/Constraints & Sources of Variation\footnotemark[1]\\
\hline
A & Spherical + $f_g$ + HSE\footnotemark[2] + $c(M)$ & None \\
B & Spherical + $f_g$ + HSE & $c$ \\
C & Spherical + $f_g$ + No Merger Boost & $c$ + Random Gas Motion \\
D & Spherical + $f_g$ & $c$ + Random Gas Motion + Merger Boost \\
E & Spherical & $c$ + Random Gas Motion + Merger Boost + $f_g$ \\
\hline
S & Simulated & All \\
SS & Simulated + Spherical & All - Morphology \\
\hline
\hline
\multicolumn{3}{l}{\footnotesize \footnotemark[1] All include scatter resulted from particle shot noise, finite cell resolution, and the simulated observation procedure.}\\
\multicolumn{3}{l}{\footnotesize \footnotemark[2] Hydrostatic equilibrium enforces no merger boost and no random gas motion.}\\
\end{tabular}
\end{center}
\label{tbl:ideal_sample}
\end{table*}

\subsection{Idealized Cluster Samples}
\label{gen_sample}

One of our main goals is to investigate the possible sources of intrinsic scatter, including the variations due to concentration, departure from hydrostatic equilibrium, merger boosts, and cluster morphology. In order to distinguish the contribution from each source, we construct a set of idealized cluster samples with different assumptions. Starting from a sample with the most possible constraints, we add one source of scatter at a time. By comparing the scatter of the idealized samples and the simulated sample, we can tell whether the scatter can be successfully reconstructed, or yet other sources still need to be found.

To this end we construct five idealized samples, going from sample $A$, with the most constraints, to sample $E$, which includes the most sources of variation. For sample $A$, only the cluster masses are taken from the simulation. Given the mass, the halo concentration $c$ is computed using the best-fit $c$--$M$ relation from \cite{2006ApJ...646..815S}. With the mass and halo concentration, the total density and gas density are assigned using the NFW profile (Navarro, Frenk, \& White \citeyear{1995MNRAS.275..720N, 1996ApJ...462..563N}, hereafter NFW) and a core-softened NFW profile \citep{2007MNRAS.380..877S}, respectively. In the core-softened NFW profile, the core radius is set to be 0.02 times the virial radius, and the gas fraction, $f_g=0.12$, is also fixed.
The pressure and temperature profiles are then computed assuming hydrostatic equilibrium (HSE). Finally, dark matter particles and zone-averaged cell quantities are assigned according to the profiles, assuming spherical symmetry. They are stored using the same file format as the simulated clusters, allowing them to be analyzed in the same way as the simulated clusters. Any scatter in the resulting $Y$--$M$ relation can only be due to particle shot noise, finite cell resolution, and the simulated observation procedure. 

Sample $B$ is generated using a similar procedure, except that the $c$--$M$ relation is assumed to have a  lognormal distribution with a dispersion of 0.22 \citep{2000ApJ...535...30J, 2001MNRAS.321..559B,2004A&A...416..853D, 2006ApJ...646..815S}. In other words, the variation in concentration should be the only additional source of $Y$--$M$ scatter for sample $B$. 

In addition to the variation in concentration, clusters in sample $C$ are allowed to depart from HSE due to pressure support from random gas motions after mergers. For each cluster, we compute the gas velocity dispersion's radial profile from the simulation and include an extra pressure term, $P_{\rm rand}=\rho \sigma^2$, in the HSE equation for computing the thermal pressure. In this way, we are effectively taking into account the incomplete virialization after merger events as another source of scatter. 

With sample $D$, we model the effect of mergers by including not only incomplete relaxation but also the merger boosts due to shock heating \citep{2001ApJ...561..621R, 2007MNRAS.380..437P}. Therefore, we extract the time histories of mass and integrated SZ flux during mergers from the ideal merger simulations in \cite{2007MNRAS.380..437P} and apply a boost in $Y$ to each idealized cluster using the actual times since last merger in the simulation and the mass ratios of those mergers. Note that the time evolution of mass from \cite{2007MNRAS.380..437P} is measured for overdensity $\Delta=500$, but the integrated SZ flux is only available for $\Delta=2500$. Thus the $Y$--$M$ scatter of sample $D$ should be considered as an upper limit to the effects of merger boosts.

Sample $E$ adds variation in the gas fraction by using the actual gas fraction computed for each simulated cluster. Sample $E$ essentially includes all possible sources of scatter investigated in this paper except the effect of cluster morphology. In \S~\ref{analyze_sample} we will compare results from these idealized samples to the simulated clusters (Sample $S$). Because the idealized samples are all constructed under the assumption of spherical symmetry, we derived another sample (Sample $SS$) using gas profiles directly extracted from the simulated clusters, such that it includes all sources of scatter except the morphological effect. These models are summarized in Table \ref{tbl:ideal_sample}. 


\section{The $Y$--$M$ Scaling Relation}
\label{Sec:result}

\begin{figure}[tbp]
\begin{center}
\includegraphics[width=0.46\textwidth]{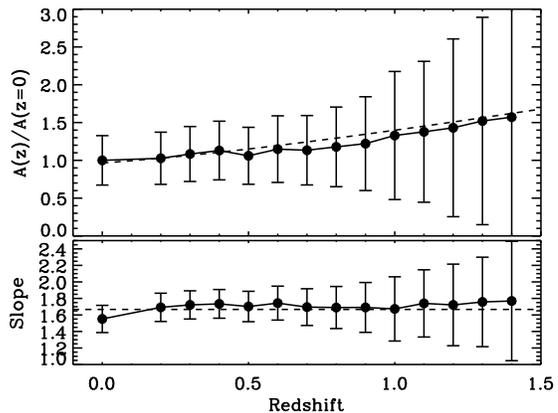}
\caption{Normalization and slope of the $Y$--$M$ relation as functions of redshift. The dashed lines are the self-similar prediction.}
\label{scaling}
\end{center}
\end{figure}

\subsection{Normalization and Slope}

For each simulation output between $z=0$ and $z=1.5$, we derive the $Y_\Delta-M_\Delta$ relation for all clusters with $M_{500} \geqslant 2\times10^{13} M_\odot$ (619 clusters at $z=0$ and 223 clusters at $z=1$) and fit it with a power law of the form
\begin{equation}
\bar{Y}_\Delta=10^{-6} A_{14}(z) \left( \frac{M_\Delta}{10^{14} h^{-1} M_\odot} \right)^\alpha,
\end{equation}
where the normalization at $10^{14}h^{-1}M_\odot$ in units of $10^{-6}$, $A_{14}$, and slope $\alpha$ are found by fitting the data points using a Levenberg-Marquardt algorithm. An overdensity of $\Delta=500$ is adopted throughout the paper. 

The normalization (relative to the value at $z=0$), slope, and scatter as functions of redshift are plotted in Figure~\ref{scaling}. As also found in previous adiabatic simulations \citep{2004MNRAS.348.1401D, 2005ApJ...623L..63M}, the evolution of the normalization and slope is consistent with the self-similar prediction, $\alpha=5/3$ and $A_{14}(z) \propto E(z)^{2/3}$, where $E(z) = [\Omega_{m0}(1+z)^3+\Omega_{\Lambda 0}]^{1/2}$, though we cannot rule out the case of no evolution because the small number of high-mass clusters at higher redshift limits the constraining power of the data. We find $A_{14}(z=0) = 5.43 \pm 0.47$, which is in agreement with the adiabatic runs in \cite{2006ApJ...650..538N} ($A_{14}(z=0) = 4.99, f_b = 0.14$) and \cite{shaw07} ($A_{14}(z=0) = 4.07, f_b = 0.11$) after taking into account the differences in the baryon fraction used in the simulations ($f_b = 0.167$ in our simulation). 


\subsection{Scatter in the $Y$--$M$ Relation}
\label{subsec:scatter}

\begin{figure}[tbp]
\begin{center}
\includegraphics[width=0.46\textwidth]{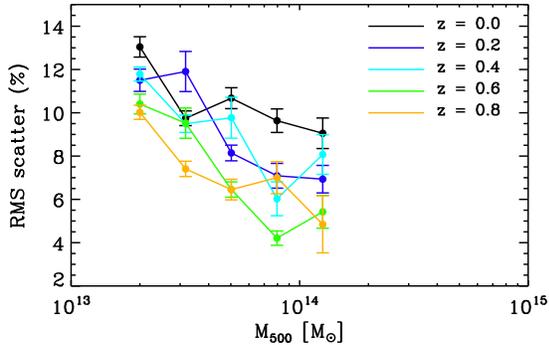} 
\caption{RMS scatter as a function of $M_{500}$. Different curves are for different redshifts. In general, the scatter decreases with both mass and redshift. See Eq.\ \ref{eq:scatter_fit} for the best-fit relation.}
\label{scatter_mbin}
\end{center}
\end{figure}

The RMS scatter around the best-fit relation is defined for $N$ clusters as
\begin{equation}
\sigma_{YM} = \left[ \frac{\Sigma_{i=1}^N (\log Y_i-\log \bar{Y}_i)^2}{N-1} \right] ^{1/2},
\end{equation}
where $Y_i$ is the measured flux of the $i$th cluster, and $\bar{Y_i}$ is the flux predicted by the best-fit relation for that cluster. Hereafter, we use the notation
\begin{equation}
\delta \log Y \equiv \log Y - \log \bar{Y}
\label{eq:dy}
\end{equation}
for the deviation from the mean relation for each cluster. For each redshift from $z=0$ to $z=1$, we compute the RMS scatter for 5 mass bins and plot the result in Figure~\ref{scatter_mbin} 
(the data for some of the redshifts are omitted for clarity).
The scatter is $\sim 5-15\%$, consistent with previous findings \citep[e.g.][]{2006ApJ...650..538N}. Moreover, we find that in general the scatter decreases with both mass and redshift. We fit the scatter using the functional form
\begin{equation}
\sigma(M,z) = A \log M + B \log(1+z) + C,
\label{eq:scatter_fit}
\end{equation} 
where the best-fit coefficients are $A = -7.06\pm0.28$, $B = -11.20\pm0.81$, and $C = 7.70\pm0.19$. The mass dependence may be due to increasing non-lognormality of the scatter when considering low-mass clusters (see \S~\ref{subsec:nonlognormal} for details). The redshift evolution may be understood by considering the self-similar model, in which all quantities for collapsed objects can be expressed in terms of the characteristic mass scale, $M_\star \propto (1+z)^{-6/(n+3)}$, where $n$ is the spectral index of the scale-free primordial power spectrum, $P(k)\propto k^n$ \citep{1986MNRAS.222..323K}. Therefore, Eq.\ \ref{eq:scatter_fit} is equivalent to the expression $\sigma = A' \log (M/M_\star) + B'$. Note that Eq.\ \ref{eq:scatter_fit} is the first attempt in the literature to quantify the scatter using a functional form of mass and redshift. This expression should be useful for future studies that require assumptions about the form of scatter.

\subsection{Non-Lognormal Scatter}
\label{subsec:nonlognormal}

Since the SZ flux is only proportional to the first power of gas density, it is sensitive to the contribution from low-density gas clumped along the line of sight to but not associated with a cluster. This is in contrast to the X-ray luminosity, which is proportional to the square of gas density. For this reason the $Y$--$M$ relation has been found to have a high-scatter tail in the distribution of its scatter \citep{2002ApJ...579...16W, santafe06}. Since any deviations from the lognormal scatter would bias cosmological constraints based on cluster counts \citep{shaw10}, we would like to examine whether the form of the log scatter can be well approximated by a Gaussian distribution, or whether generalizations of the parametrization need to be considered.

A purely Gaussian distribution can be described exactly using only its mean value $\mu$ and variance $\sigma^2$:
\begin{equation}
G(x) = \frac{1}{\sigma\sqrt{2\pi}}\exp\left[-\frac{(x-\mu)^2}{2\sigma^2}\right].
\label{eq:gaussian}
\end{equation}
A nearly-Gaussian distribution can be approximated using the Edgeworth expansion \citep[e.g.][]{1995ApJ...443..479B,1998A&AS..130..193B},
\begin{equation}
\tilde{G}(x) \approx G(x) - \frac{\gamma}{6} \frac{d^3 G}{dx^3} + \frac{\kappa}{24} \frac{d^4 G}{dx^4} + 
\frac{\gamma^2}{72} \frac{d^6 G}{dx^6},
\label{eq:ng}
\end{equation}
which is parametrized by four moments -- the mean and the variance describing the Gaussian distribution, plus the skewness ($\gamma$) and the kurtosis ($\kappa$) describing the deviation from gaussianity. The skewness is defined as
\begin{equation}
\gamma = \frac{\langle(x-\mu)^3\rangle}{\sigma^3}
\label{eq:skew}
\end{equation}
and the kurtosis as
\begin{equation}
\kappa = \frac{\langle(x-\mu)^4\rangle}{\sigma^4}-3.
\label{kurt}
\end{equation}

\begin{figure}[tbp]
\begin{center}
\includegraphics[width=0.4\textwidth]{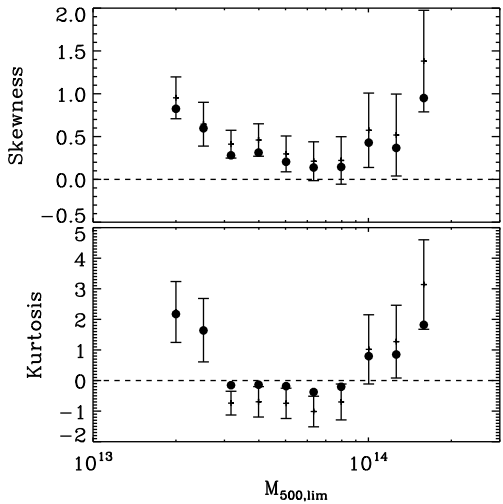}
\caption{Skewness and kurtosis of the $Y$--$M$ scatter for simulated clusters at $z=0$ as functions of limiting mass. See the text for the definition of error bars.}
\label{sk_mlim}
\end{center}
\end{figure}

We compute the skewness and kurtosis of the $Y$--$M$ scatter for clusters at $z=0$ above different mass thresholds  and plot the results in Figure~\ref{sk_mlim}. The error bars represent the uncertainty due to finite sample size and are estimated using $10^3$ Monte-Carlo realizations of random sampling from a nearly-Gaussian distribution given the measured skewness and kurtosis as in Eq.\ \ref{eq:ng}.
Note that because of finite sample size the measured skewness and kurtosis would underestimate the intrinsic values of the underlying distribution. This bias is represented by the offset between the data points and the middle points of the error bars.

\begin{figure}[tbp]
\begin{center}
\includegraphics[width=0.4\textwidth]{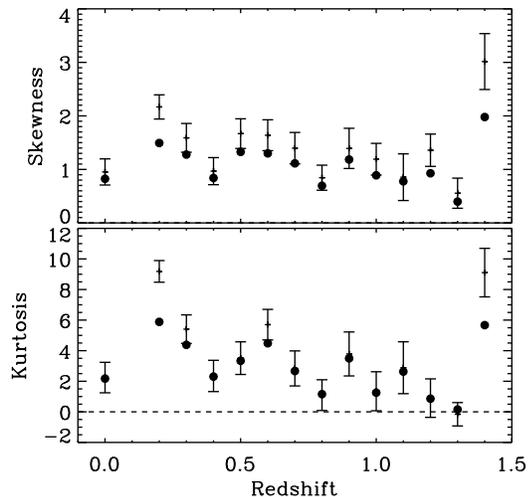}
\caption{Skewness and kurtosis of the $Y$--$M$ scatter for simulated clusters with $M_{500} \geqslant 2\times 10^{13} M_\odot$ as functions of redshift. See the text for the definition of error bars.}
\label{sk_evol}
\end{center}
\end{figure}

We find that the scatter is non-lognormal with positive skewness and kurtosis when including only massive clusters ($M_{500,{\rm lim}} \gtrsim 10^{14}M_\odot$) or when more and more low-mass clusters are included ($M_{500,{\rm lim}} \lesssim 5 \times 10^{13}M_\odot$). Due to the limited number of clusters at the high-mass end, the log scatter there is expected to follow Poisson statistics and deviate from a Gaussian form. For the lower mass range, on the other hand, we find that there is a tail toward positive values in the distribution of scatter, which increases the skewness and kurtosis. By visual inspection of clusters in the tail, we find that these objects happen to have elongated shapes or clumped gas along the line of sight. Since less massive clusters are more likely to be surrounded by gas with mass comparable to their own, this effect becomes more important for lower-mass clusters. We will further address this point in \S~\ref{subsec:morphology}. For other redshifts, the level of non-lognormality is also non-negligible, as shown in Figure~\ref{sk_evol}. The median skewness and kurtosis are 1.43 and 4.21, respectively.

\section{Sources of Scatter}
\label{Sec:source}

We now investigate the physical origin of the intrinsic scatter for the purpose of understanding the above trends and reducing it for better mass estimates. Possible sources of scatter in our simulation include halo concentration, dynamical state, and cluster morphology. We examine each effect by correlating the scatter with each individual source for clusters at $z=0$. We show that the scatter can be reduced by choosing appropriate measures of each effect. At the end of this section we compare the percentage contribution from each source using the idealized cluster samples described in \S~\ref{gen_sample}. 

\subsection{Concentration}
\label{subsec:cc}

\begin{figure*}[thbp]
\epsscale{1.0}\plottwo{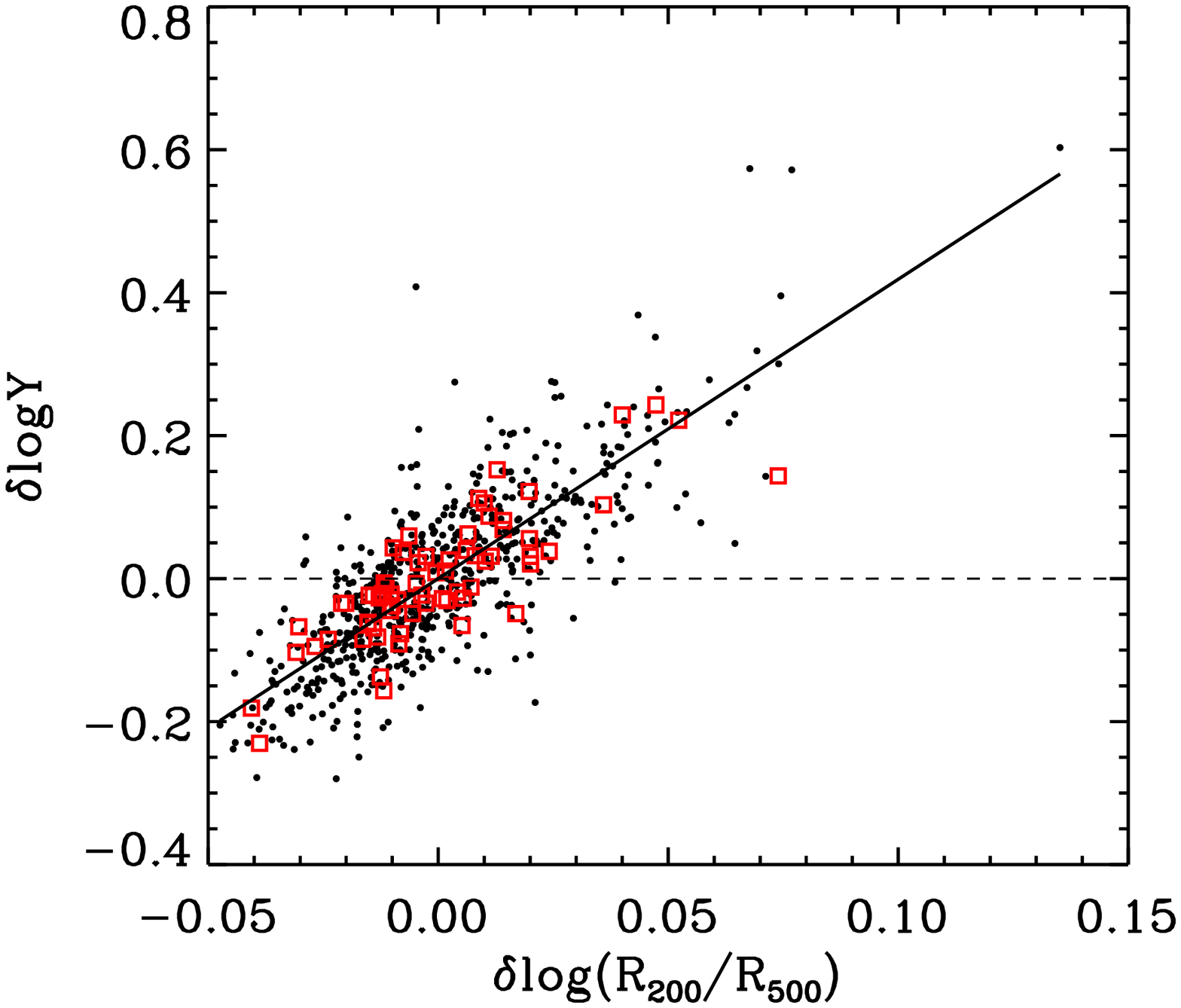}{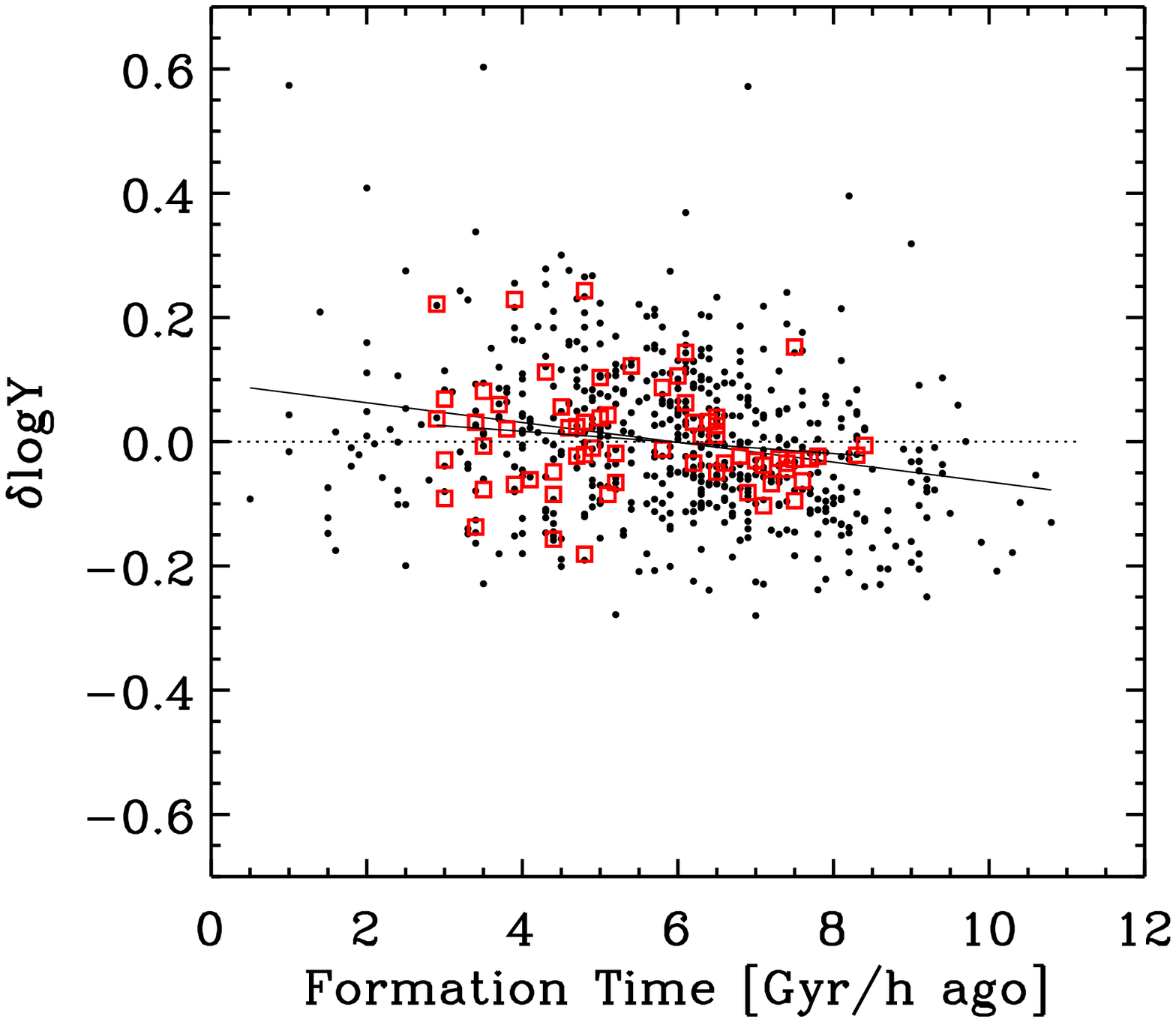}
\caption{{\em Left}: $Y$--$M$ scatter versus $(R_{200}/R_{500})$--$M$ scatter at $z=0$, where $R_{200}/R_{500}$ is a monotonically decreasing function of halo concentration. Correlation coefficient is 0.775.
{\em Right}: $Y$--$M$ scatter versus formation time at $z=0$. Correlation coefficient is -0.244. Clusters with $M_{500} \geqslant 10^{14}M_\odot$ are plotted using open squares.}
\label{dy_dc_tf}
\end{figure*}

The concentration of a dark matter halo is usually defined as the ratio between the virial radius and the NFW scale radius, i.e., $c \equiv R_{\rm vir}/R_s$. The concentration parameter characterizes the density inside the core region of a halo and reflects the mean density of the universe when the halo collapsed. Thus halos formed earlier in time tend to be more concentrated (NFW \citeyear{1997ApJ...490..493N}; \citeauthor{2002ApJ...568...52W} 2002). By testing the correlation of $Y$--$M$ scatter with scatter in the concentration parameter, we are effectively probing the influence of cluster formation history.

We choose to use the parameter $R_{200}/R_{500}$ instead of the original halo concentration parameter $c$ because it has two advantages. The first is that it avoids introducing the uncertainty of fitting an NFW profile, especially for less massive clusters, since the fitting is very sensitive to the grid resolution in the central region of a cluster. Moreover, our analyses involve not only relaxed clusters but also merging ones, for which $R_{200}/R_{500}$ is actually better-defined than $c$, since an NFW profile would yield a poor fit.

Figure~\ref{dy_dc_tf} (left panel) shows a strong positive correlation between scatter in 
the $Y$--$M$ relation and scatter in the $(R_{200}/R_{500})$--$M_{500}$ relation.
The correlation coefficient is 0.64, with a probability of zero given by the Spearman Rank-Order Correlation test (\citeauthor{1992NumericalRecipe}\ 1992, \S14.6; probability of one means no correlation). To ensure that this result is not biased by the lower-mass clusters whose $R_{500}$ values are close to the resolution of the simulation, we raised the mass threshold to $M_{500} \geqslant 10^{14} M_\odot$ and found that the result is robust for these well-resolved systems (shown as the open squares in Figure~\ref{dy_dc_tf}). 
Note that we correlate with $\delta \log(R_{200}/R_{500})$ instead of the raw value of $R_{200}/R_{500}$ because the latter is a function of cluster mass. By doing so we exclude the effect of
different cluster masses, focusing on the variation in halo concentrations.
$R_{200}/R_{500}$ is a monotonically {\it decreasing} function of the halo concentration parameter (see \cite{2009ApJ...699..315Y} for derivation). Therefore, for clusters with similar masses, more concentrated clusters tend to lie under the mean $Y$--$M$ relation, while the ``puffier'' clusters tend to scatter high.

Since halo concentration is related to cluster formation time, we can test the above trend by checking the correlation of $Y$--$M$ scatter with the formation times derived from the mass assembly histories of our simulated clusters. The formation time here is defined as the time when the cluster first exceeds half of its final mass. As expected, we find that clusters that formed earlier (thus with higher concentrations) tend to scatter low (see right panel of Figure~\ref{dy_dc_tf}). The correlation is not as tight as the one with the halo concentrations. This is due to the fact that the correlation between the halo concentration and the cluster formation time itself has a very large scatter, and also that the variation in halo concentrations cannot be fully accounted for by the variation in cluster formation time \citep{2007MNRAS.381.1450N}. But the direction of the correlation with cluster formation time is consistent with the correlation with halo concentration. 

To explain the correlation between the $Y$--$M$ scatter and the concentration, recall the virial theorem for the simplest case of an isolated system: $2K+U=0$, where $K$ and $U$ are the total kinetic and gravitational binding energies of the system, respectively. In general one can write
\begin{equation}
\frac{k_B T}{\mu m_p} \propto \frac{GM}{R},
\label{eq:tm}
\end{equation}
where $\mu$ is the mean molecular mass of the gas, $m_p$ is the mass of a proton, and $T$, $M$, and $R$ are the virial temperature, mass, and radius of the system,
respectively. Together with the definitions of the SZ flux (Eq.\ \ref{eq:yint}) and $M={4\over 3}\pi R^3 \bar{\rho}$, one can derive 
\begin{equation}
Y \propto M_{gas} T \propto f_g M^{5/3}.
\label{eq:ym}
\end{equation}
Note that the above relations are for virial quantities of a cluster {\it as a whole}, but mass-observable relations are often measured using a certain aperture size, $R_\Delta$. The relation between the virial and overdensity quantities depends on individual cluster profiles, which are determined by the halo concentration and how the gas is distributed on top of the dark matter potential (e.g.\ equation of state of gas). Therefore, the normalization, and thus the scatter, of the $Y_\Delta$--$M_\Delta$ relation is a function of halo concentration and gas properties. 

\begin{figure*}[thbp]
\epsscale{1.0}\plottwo{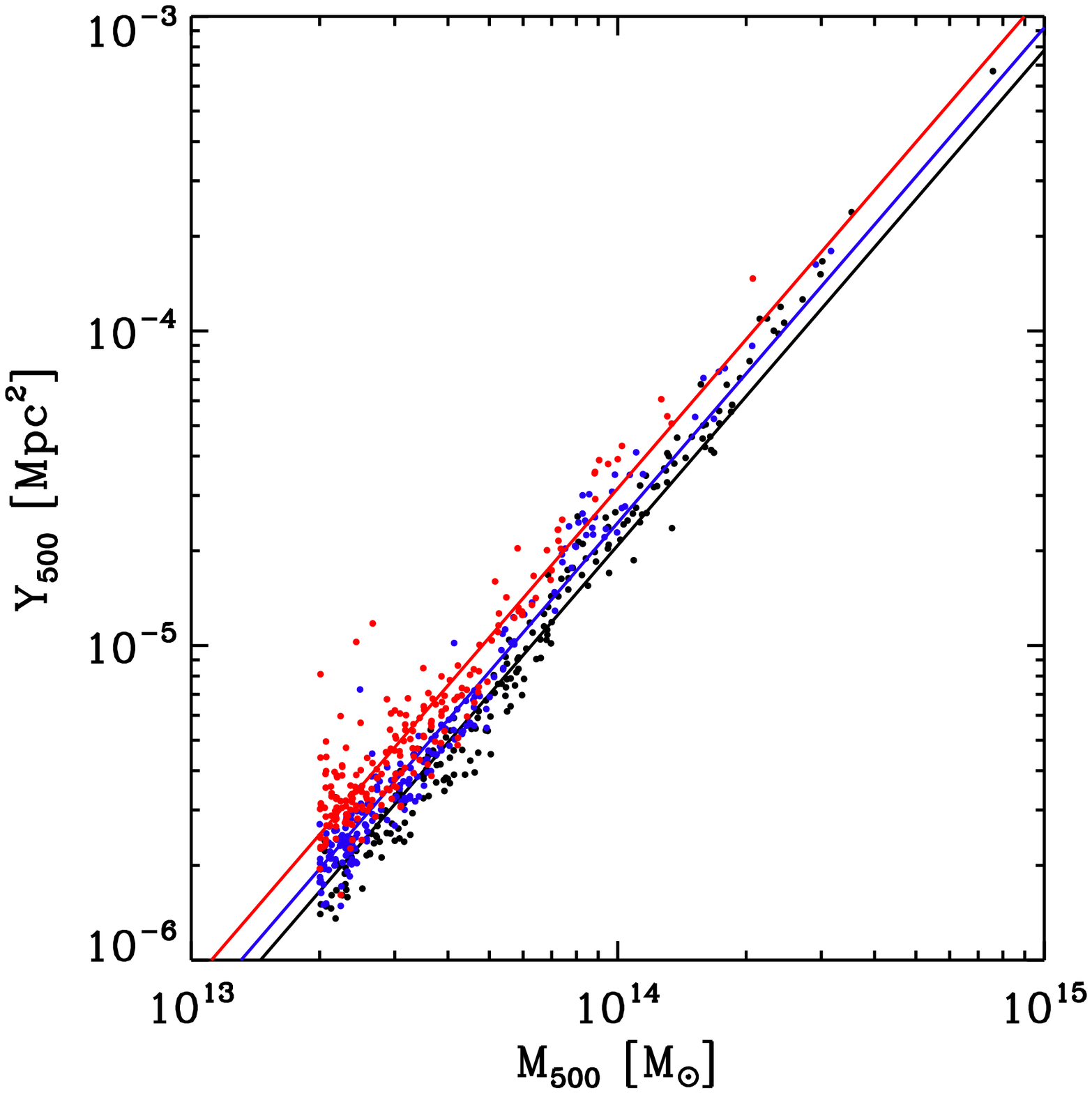}{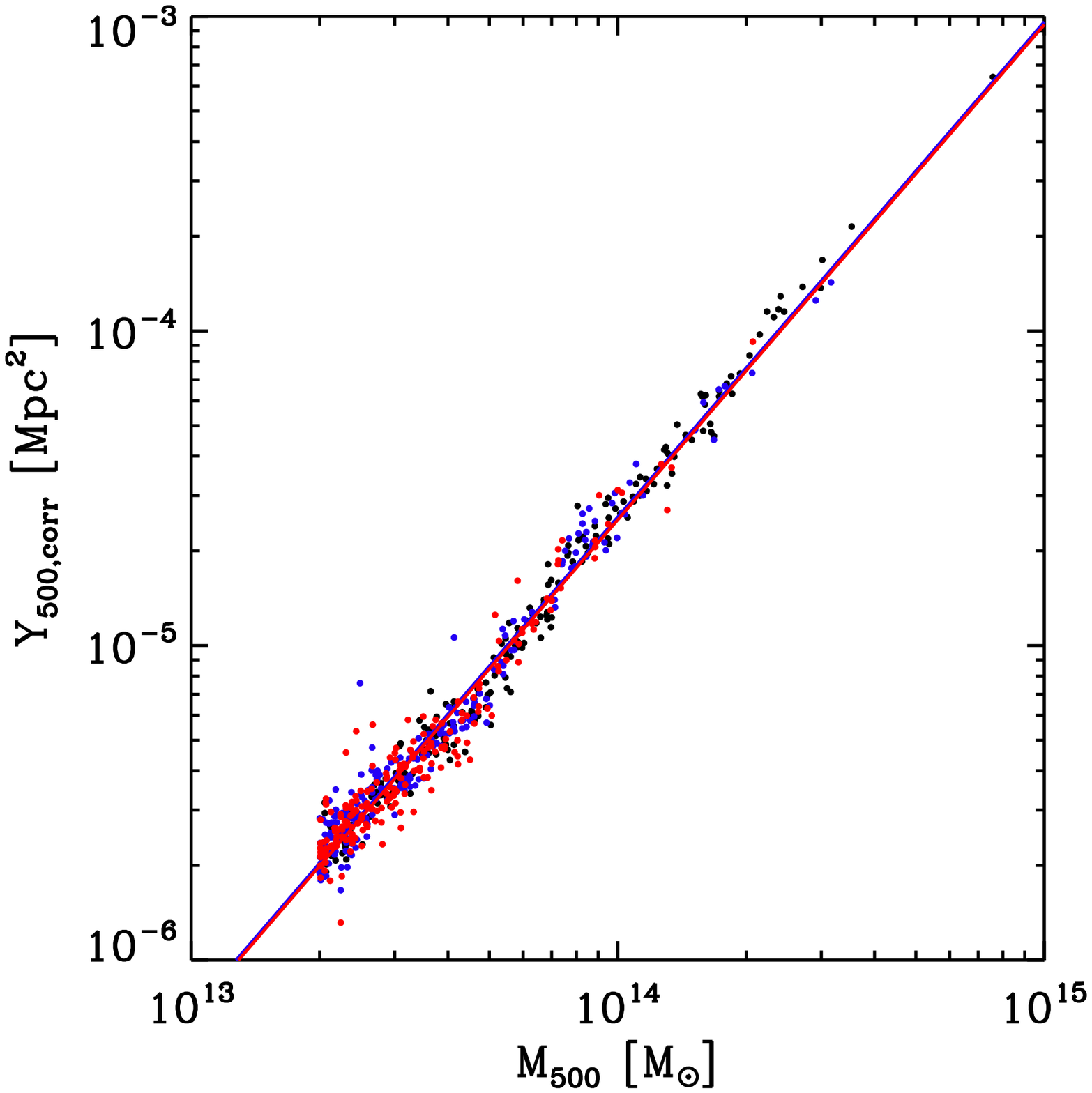}
\caption{{\em Left}: The $Y$--$M$ relation plotted for clusters having different values of concentration at $z=0$. The 1/3 percentiles with the highest, intermediate, and lowest values of concentration are plotted 
using black, blue, and red symbols, respectively.  
{\em Right}: The $Y$--$M$ relation after correction for the dependence on concentration using Eq.\ \ref{eq:correction}.}
\label{myc}
\end{figure*}

It is also important to note that, from the above derivation, the direction of the correlation between the scatter and concentration is dependent on the gas properties. Our result shows that less concentrated clusters tend to scatter high, i.e., have higher pressure than clusters of similar masses. 
Observationally, \cite{2010ApJ...715..162C} has also found a similar anti-correlation between the X-ray temperature-mass scatter and strong lensing concentration.
This may be attributed to the fact that less concentrated clusters have larger scale radii, and hence when comparing with clusters of the same $M_\Delta$ or the same aperture size $R_\Delta$, their ratios $R_\Delta/R_s$ are smaller, which means the observable integrated within $R_\Delta$ would be greater. However, different simulations can have different directions of correlation depending on the input gas physics. For example, \cite{shaw07} also found a correlation between the scatter and concentration, but in the opposite direction. The difference may be due to different gas physics included in their models.
In principle, by assuming a particular gas model the constant of proportionality in Eq.\ \ref{eq:ym} can be computed exactly. \cite{2006MNRAS.371..193A} has done this exercise assuming a polytropic equation of state for the gas. According to their calculation, the coefficient in the $T$--$M$ relation (as in Eq.\ \ref{eq:tm}; inverse of the $Y_{MT}$ in their Eq.\ 17) decreases with concentration for a polytropic index of $\gamma_p=5/3$. As $\gamma_p$ decreases, the dependence becomes weaker and then the direction is reversed. Since including extra baryonic physics effectively works to decrease $\gamma_p$ (e.g. {\it for fixed mass and concentration}, both changes yield a shallower temperature profile, see also Figures 2 and 3 in \cite{ostriker06}), this may explain why the dependence on concentration can be different between models with different input gas physics. Note, however, that in reality the situation can be even more complicated because a constant $\gamma_p$ may not be valid for all gas models \citep{2004MNRAS.355.1091K}.

The strong correlation in Figure~\ref{dy_dc_tf} suggests that the variation in halo concentrations contributes a significant amount of the intrinsic scatter in the $Y$--$M$ relation. Using this strong correlation it is possible to adjust for the dependence of SZ flux on cluster concentrations. We use $\delta \log(R_{200}/R_{500})$ for each cluster to calculate its expected $\delta \log Y$ from the best-fit relation, $(\delta \log Y)_{\rm exp} = 7.167 \times \delta \log(R_{200}/R_{500})$. We then subtract this $(\delta \log Y)_{\rm exp}$ from the measured SZ flux to obtain a corrected flux, 
\begin{equation}
(\delta \log Y)_{\rm corr} = \delta \log Y - (\delta \log Y)_{\rm exp}.
\label{eq:correction}
\end{equation}
The $Y$--$M$ relations before and after correcting for concentration are shown in Figure~\ref{myc}. We find that after removing the effect of halo concentration, the RMS scatter decreases from 12.07\% to 7.34\% (i.e.\ by 38.9\%). 
This method was proposed by \cite{2009ApJ...699..315Y} to tighten the X-ray temperature-mass relation and has been successfully applied to observed strong lensing clusters \citep{2010ApJ...715..162C}. In addition to strong lensing, the NFW concentration can also be measured via weak lensing, X-ray emission, etc.\ \citep[][and references therein]{2007MNRAS.379..190C, 2008JCAP...08..006M, 2007ApJ...664..123B}, although one has to be careful about systematics of each method and when combining different measurements.
In fact it can be generalized to any observable other than concentration. That is, if there exists any variable $X$ whose mass scatter $\delta \log X$ is known and correlates with $\delta \log Y$, then given the best-fit correlation $(\delta \log Y)_{\rm exp} = \alpha \times \delta \log X$, the $Y$--$M$ scatter can be reduced in a similar way using Eq.\ \ref{eq:correction} to remove the effect of $X$ from the scatter. Therefore, this method can be a powerful way to reduce the observed mass-observable scatter and obtain better mass estimates.


\subsection{Dynamical State}

Another possible origin of the $Y$--$M$ scatter is cluster dynamical state. Cluster mergers are among the most energetic events in the universe. Shock heating and departure from hydrostatic equilibrium during mergers can drive clusters away from the mean scaling relations. Ideal merger simulations \citep{2001ApJ...561..621R, 2007MNRAS.380..437P} have shown that the effect of shock heating to boost the SZ and X-ray observables to values a few times higher than the pre-merger values. Studies that combine the amount of boosting predicted by these simulations with extended Press-Schechter merger trees \citep{2002ApJ...577..579R, 2008ApJ...680...17W} show that the boosting effect can bias estimates of cosmological parameters such as $\sigma_8$ and $\Omega_m$. The other effect of mergers is departure from hydrostatic equilibrium. Before the gas within a merger is completely virialized, the pressure support from random gas motions can contribute $\sim 10-20\%$ of its thermal pressure \citep{2006MNRAS.369.2013R, 2009ApJ...705.1129L}. Thus the thermal pressure and hence the SZ flux of unrelaxed systems is expected to be smaller than that of relaxed systems of similar masses. These previous studies are primarily based on small cluster samples. Therefore, our aim is to investigate how merger events {\it statistically} influence the cluster scaling relations.   

\begin{figure*}[thbp]
\epsscale{1.0}\plottwo{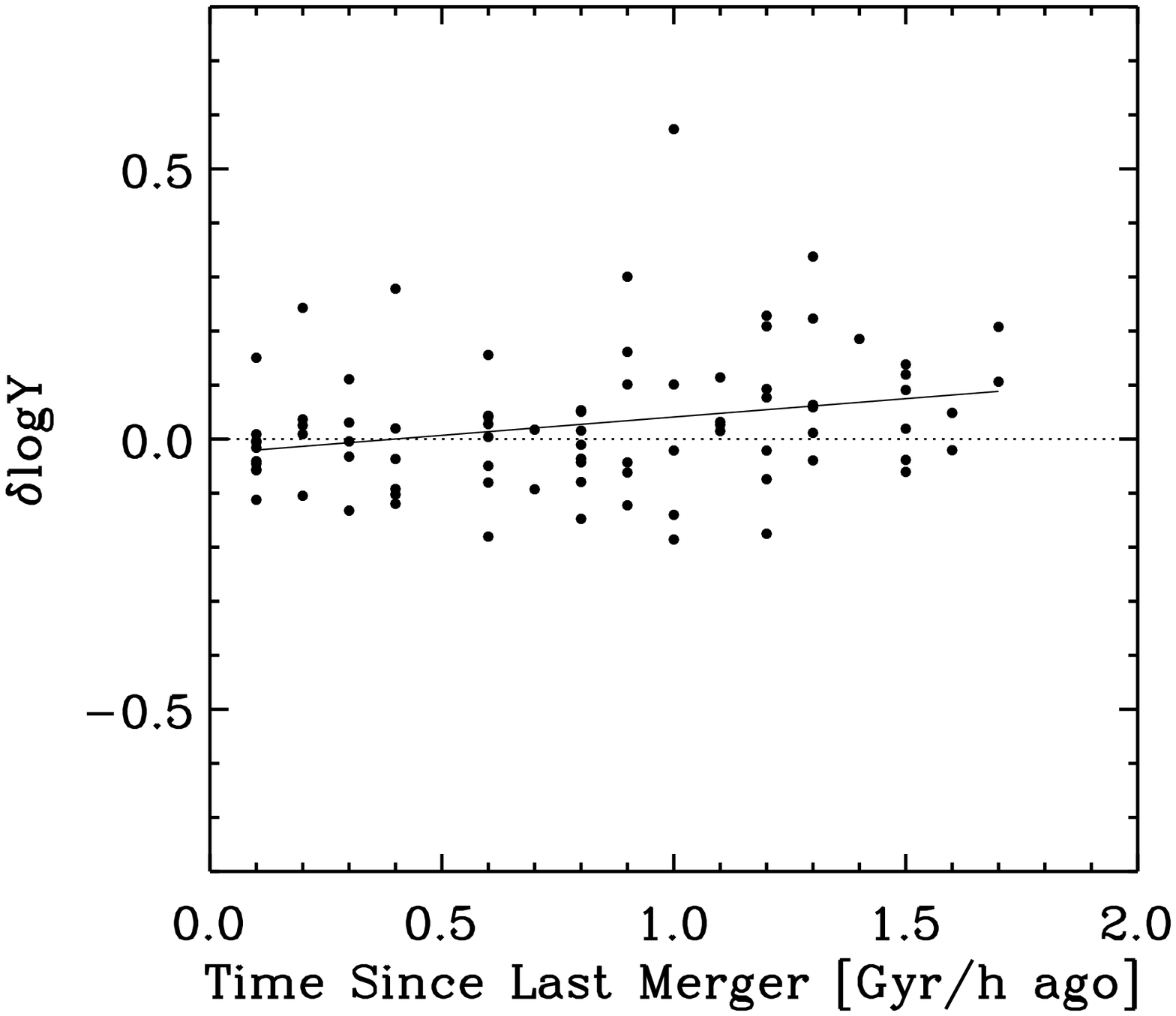}{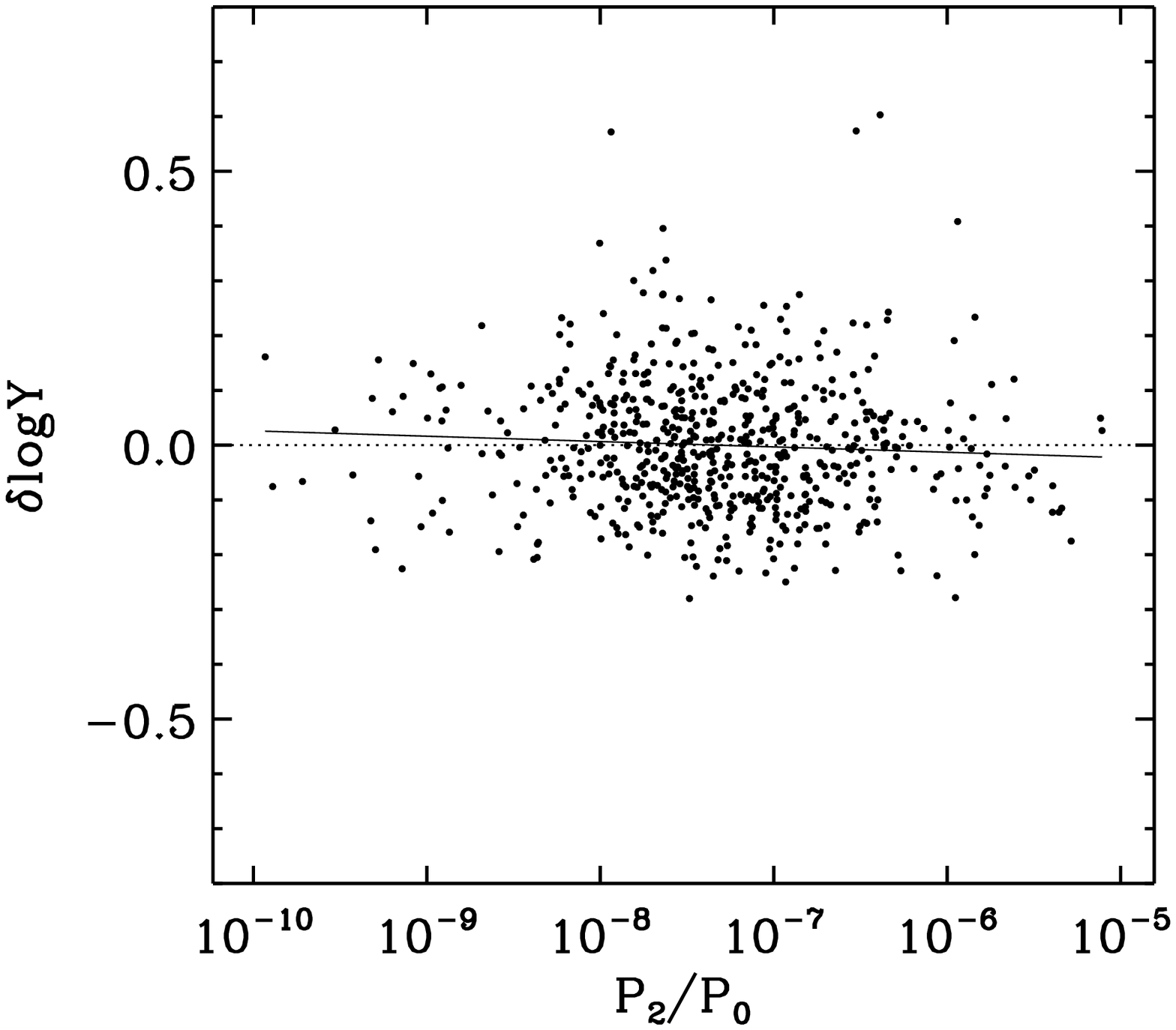}
\caption{{\em Left}: $Y$--$M$ scatter versus time since last merger for major mergers at $z=0$. The correlation is statistically significant but weak (correlation coefficient of 0.299; probability of no correlation of 0.006). 
{\em Right}: $Y$--$M$ scatter versus one of the power ratios, $P_2/P_0$. Correlation coefficient is -0.091; probability of no correlation is 0.023.}
\label{fig:dy_dyn}
\end{figure*}

If the scatter were dominated by the boosting effect of cluster mergers as described above, then one would expect to find merging clusters to preferentially lie above the mean relation. However, if during mergers the departure from hydrostatic equilibrium due to non-thermal pressure support were dominant, mergers would tend to scatter low. In order to see which effect is more prominent, we correlate the $Y$--$M$ scatter with the time since last merger (Figure~\ref{fig:dy_dyn}, left panel). Substructure measures such as centroid offset \citep{1995ApJ...447....8M} and power ratios \citep{1995ApJ...452..522B,1996ApJ...458...27B} are often used to quantify departures from equilibrium in clusters. We compute the centroid offset and power ratios for the simulated clusters using the same definition as in \cite{2009ApJ...699..315Y}. The right panel of Figure~\ref{fig:dy_dyn} shows the correlation with one of the power ratios, $P_2/P_0$. Based on the Spearman Rank-Order Correlation test (\citeauthor{1992NumericalRecipe}\ 1992, \S14.6), both correlations are weak (with a small correlation coefficient) but significant (with a high probability), in the direction that more disturbed clusters tend to scatter low. This implies that during mergers the incomplete virialization may be the more important factor in driving the scatter than the shock heating effect. However, the fact that these two effects operate in opposite directions may be the reason why there is not a clear trend with merger activities. Moreover, a number of factors can dilute the shock boosting effect, such as the small chance of finding mergers in progress, capturing the shocks within an overdensity radius at the right projection and at the right moment during a merger's transient boost, and the fact that merging clusters tend to move along the scaling relations because their masses also increase at the same time their observables increase, as found also by \cite{2008ApJ...680...17W} and \cite{2006ApJ...650..128K} (see \cite{2009ApJ...699..315Y} for an extensive discussion).

\begin{figure}[thbp]
\begin{center}
\includegraphics[width=0.45\textwidth]{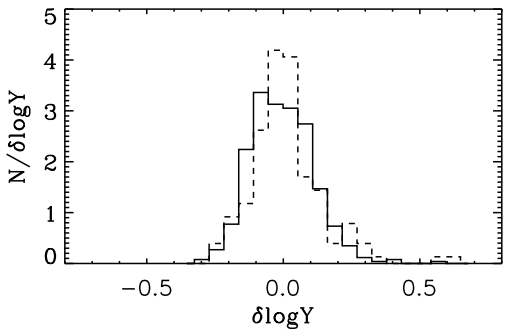}
\caption{Normalized distribution of the $Y$--$M$ scatter for relaxed (solid) and merging (dashed) clusters at $z=0$. According to results from significance tests, mergers do not have a bias but do have a larger dispersion with respect to the relaxed clusters.}
\label{fig:dydist}
\end{center}
\end{figure}

Do mergers bias the $Y$--$M$ relation due to incomplete virialization? To answer this question, we plot the normalized distributions of the $Y$--$M$ scatter for relaxed and merging clusters in Figure~\ref{fig:dydist} and use the Wilcoxon Rank-Sum (R-S) test and the F-variance (F-V) test to see whether these two populations have significantly different mean values or variances, respectively. A value smaller than 0.05 (for a significance level of 5\%) returned by these tests is commonly adopted to indicate a significant difference between two populations. We find that their mean values do not differ significantly, but mergers have a wider distribution compared to relaxed clusters (with significance 0.014). Therefore, although mergers do not tend to bias the scaling relation, they do have a greater amount of scatter than relaxed clusters. If merging (relaxed) clusters are chosen to be those within the quartile with the highest (lowest) substructure measures, similar trends with significant probabilities are found for 9 out of 21 substructure measures ($P_2/P_0$, $P_3/P_0$, and the centroid offset $W$ measured from different viewing directions and varying aperture sizes). We find that including only the merging clusters would result in $\sim 15-45\%$ greater scatter than when only relaxed clusters are taken into account, consistent with previous findings \citep{shaw07}. Note however that separating mergers from relaxed clusters does not reduce the skewness or kurtosis of the scatter distribution, which suggests that the non-lognormality has causes other than mergers (see \S~\ref{subsec:morphology}). 


\subsection{Morphology}
\label{subsec:morphology}

Despite the spherical symmetry that theoretical models usually assume, both simulations \citep{1992ApJ...399..405W, 1996MNRAS.281..716C, 2002ApJ...574..538J, 2005ApJ...627..647B, 2005ApJ...629..781K} and observations \citep[e.g.][]{2000MNRAS.316..779B} have shown that clusters are triaxial (or elliptical when projected) rather than simple spheres, even for relaxed clusters. How the gas is distributed in the cluster potential well should vary depending on the axes ratios of the cluster. Moreover, viewing a triaxial cluster from different angles should also yield different observed quantities integrated along the line of sight. Both these factors can contribute to the $Y$--$M$ scatter.  

In order to explore the impact of morphology, for each simulated cluster we find the orientation of the principal axes by diagonalizing the moment of inertia tensor, $I_{\alpha\beta}=m_i \Sigma_i r^\alpha_i r^\beta_i$, where the summation is over all the particles and cells in the cluster, $m_i$ is the mass of a particle or a gas cell, and $r^\alpha_i$ is the $x$, $y$, or $z$ component of the distance from the cluster center of mass. The lengths of the major, intermediate, and minor axes, denoted as $a$, $b$, and $c$, are found by finding the intercepts of the axes with the isodensity surface having overdensity $\Delta=200$. The angles $\theta_\alpha$ are defined to be the angles between the major axis and the directions of projection ($\alpha=x,y,z$; note that the $x$-direction is the projection used for all the analyses in this paper). 

\begin{figure}[tbp]
\begin{center}
\includegraphics[width=0.4\textwidth]{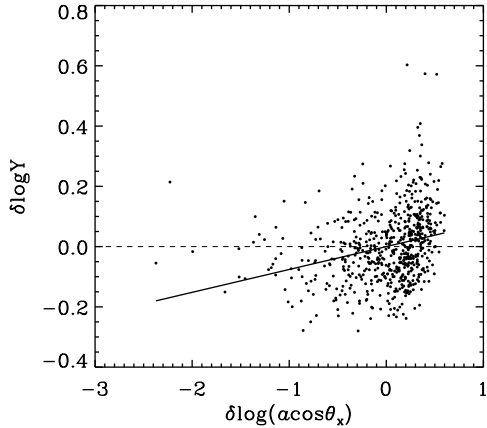}
\caption{Correlation between the $Y$--$M$ scatter and the $a\cos \theta_x$--$M$ scatter at $z=0$. Clusters that are more elongated along the line-of-sight have larger values of $a\cos \theta_x$. Correlation coefficient is 0.345.} 
\label{fig:dy_dacos}
\end{center}
\end{figure}

Motivated by our results in \S~\ref{subsec:nonlognormal} that the non-lognormality may be due to clusters that happen to have elongated shapes aligned with the viewing direction, we invented a measure, $a\cos\theta_x$, to trace cluster morphology along the line of sight. Figure~\ref{fig:dy_dacos} shows the $Y$--$M$ scatter versus the scatter in the $a\cos\theta_x$--$M$ relation. The positive correlation indicates that clusters that are more elongated along the line of sight preferentially have higher $Y$--$M$ scatter, which is expected because the SZ flux is roughly proportional to the column density of the gas (see Eq.\ \ref{eq:ylos}). Given the best-fit relation, $\delta \log Y = 0.076 \times \delta \log (a\cos \theta_x)$, we can again reduce the scatter by applying a correction as in Eq.\ \ref{eq:correction}. By doing so we find that the scatter is reduced from 12.07\% to 11.63\% (i.e.\ by $3.6\%$).  

\begin{figure}[tbp]
\begin{center}
\includegraphics[width=0.45\textwidth]{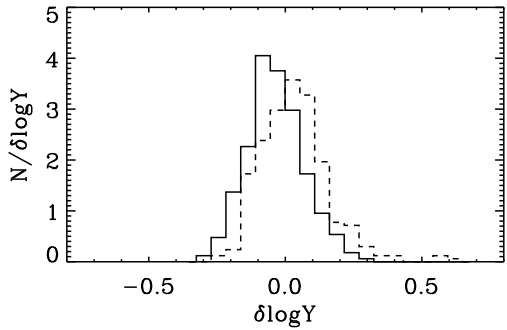}
\caption{Normalized distributions of the $Y$--$M$ scatter for clusters with the morphology measure $a\cos \theta_x$ smaller than the median (solid) and larger than the median (dashed). The scatter distribution for clusters with elongated shape along the line-of-sight (dashed; $\gamma=1.04$, $\kappa=2.63$) is much more non-lognormal than that of the remaining population (solid; $\gamma=0.29$, $\kappa=-0.08$).} 
\label{fig:dydist_acos}
\end{center}
\end{figure}

We further divide the cluster sample in half using the values of $a\cos\theta_x$ and plot the normalized distributions of the $Y$--$M$ scatter in Figure~\ref{fig:dydist_acos}. We find that the scatter distribution of the clusters with larger $a\cos\theta_x$ is non-lognormal ($\gamma=1.04$, $\kappa=2.63$), while the skewness and kurtosis of the remaining population are greatly reduced ($\gamma=0.29$, $\kappa=-0.08$). Therefore, the non-lognormality is indeed caused by the clusters with more elongated shapes along the line of sight.


\subsection{Projection Effects Due to Large-Scale Structure}

Since the SZ flux is obtained by integrating along the line of sight within a projected radius, it is subject to contamination by gas that lies along the same line of sight, which causes the large number of high-scatter objects in the $Y$--$M$ relation found in simulations that include light cones \citep{2002ApJ...579...16W, santafe06}. These outliers and the outliers due to morphology discussed in the previous section can both drive the non-lognormality of the scatter. Since our simulated observations only include isolated clusters and thus do not take the projection effect into account, the skewness and kurtosis estimated from our simulation may be considered to be underestimates of the true values. In this case, it is even more important to adopt the higher moments in the parametrization of the scatter in order to get unbiased cosmological constraints.


\subsection{Insights from Idealized Samples}
\label{analyze_sample}

\begin{figure}[tbp]
\begin{center}
\includegraphics[width=0.4\textwidth]{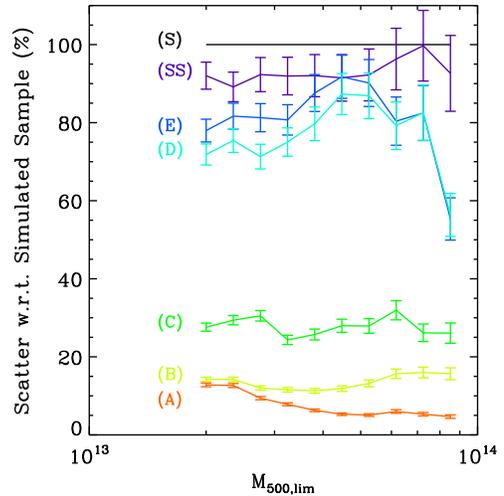}
\caption{The percentage contribution of the scatter for each idealized sample with respect to the simulated sample. The bottom (top) sample includes the least (most) physical sources of variations. See Table \ref{tbl:ideal_sample} for a summary of notations and assumptions used to construct each sample.}
\label{dy_mlim_percent}
\end{center}
\end{figure}

Figure~\ref{dy_mlim_percent} shows the percentage contribution of scatter for each idealized cluster sample with respect to the simulated sample. The cluster sample at the bottom has the most constraints on and least freedom in the model parameters, and the assumptions are loosened one at a time from bottom to top (see Table \ref{tbl:ideal_sample} for a summary of model descriptions). In general the values are independent of mass, that is, the processes shape the scaling relation in a self-similar way, as expected in the absence of additional baryonic physics. The only exception is the bottom curve for which only the masses of clusters are assigned. In principle this sample should have zero scatter if the resolution of gas cells were infinite, but in reality the finite resolution introduces a nonzero scatter which becomes bigger as more low-mass clusters are included. Because the idealized samples are all constructed under the assumption of spherical symmetry, we derived Sample $SS$ (second line from the top in the figure) using gas profiles directly extracted from the simulated clusters, such that it includes all sources of scatter except the morphological effect. In other words, the difference between the simulated clusters and the spherically-smoothed clusters is solely due to the variation in cluster morphology, which is $\sim 10\%$. 

From the differences between the subsequent samples we are able to isolate the contribution of each effect to the total scatter: the variation in halo concentration contributes $\sim 10-20\ \%$ (difference between $A$ and $B$), the departure from hydrostatic equilibrium results in $\sim 10-15\ \%$ (between $B$ and $C$), merger boosts add another $\sim 30-60\ \%$ (between $C$ and $D$), the variation in gas fractions introduces $\sim 0-10\%$ (between $D$ and $E$), and the rest (between $D$ and $SS$) due to other unaccounted-for effects is $\sim 0-30\ \%$. 
Note that the contribution from variation in concentration quoted here is estimated using sample $B$, which assumes spherical symmetry and hydrostatic equilibrium. However, in reality, both changing cluster morphology and including random gas motions \citep{2009ApJ...705.1129L} would further modify the distribution of gas and hence alter the measured value of concentration. In other words, the scatter is driven not only by the variation of concentration in sample $B$, but also by that due to cluster morphology and departure from HSE. Therefore, the total effect of concentration, as suggested by the correction for concentration in \S~\ref{subsec:cc} (i.e.\ $\sim 40\%$), would be more appropriately accounted for by also considering the contributions from morphology and random gas motions.


\section{Combining X-ray and SZ Scaling Relations}
\label{Sec:xray}

In the previous sections we have discussed various sources of intrinsic scatter in the relation between the SZ flux and the {\it true} mass. However, observationally cluster masses still need to be measured in some way, such as via X-ray hydrostatic assumptions or optical richness. Cross-calibration across measurements at different wavelengths is important because it provides a consistency check that can minimize the possible systematic effects of each individual measurement \citep[e.g.][]{sptprf}, such as the projection effects to which SZ and optical observations are subjected to. Therefore, high-precision cluster cosmology requires that we combine SZ cluster surveys with X-ray or optical follow-ups \citep{high10,act10}.

However, one needs to be cautious when combining multiple mass proxies because their errors may be correlated. For example, because both the SZ and optical signals are subject to projection effects, clusters can have consistent mass estimates that are both actually biased with respect to the true mass    \citep{2009MNRAS.393..393C}. Since it is impossible for observations to disentangle such correlations, one has to rely on numerical simulations to determine whether these effects are serious for any given pair of mass proxies. Here we would like to explore whether this correlated error exists between the SZ flux and the low-scatter X-ray mass proxy \citep{2006ApJ...650..128K}, $Y_X$, which is commonly used as a mass proxy in X-ray observations.  
Note that although individual X-ray properties such as $M_{gas}$ and $T_X$ would be affected by core properties, Stanek et al. (2007) found that the $Y_X$ parameter, which combines the effects of $M_{gas}$ and $T_X$, is remarkably insensitive to baryonic physics. That is, for their runs with and without the preheating prescription, both the amount and shape of the $Y_X$--$M$ scatter are almost identical. Therefore, our results below should be robust to additional baryonic physics.

\begin{figure}[tbp]
\begin{center}
\includegraphics[width=0.45\textwidth]{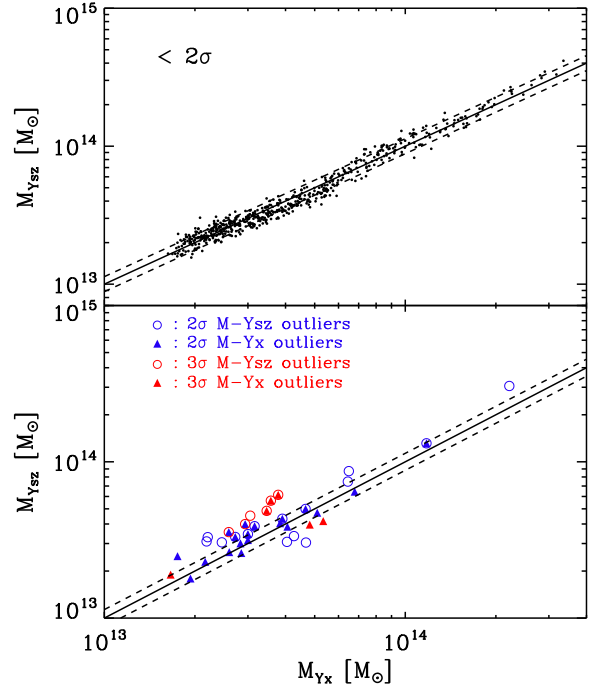}
\caption{$Y_{SZ}$ predicted mass versus $Y_X$ predicted mass. Clusters in the {\em upper} panel have less than $2\sigma$ deviations from both the mean $Y_{SZ}$--$M$ relation and $Y_X$--$M$ relation, while clusters whose mass scatter is bigger than $2\sigma$ for either relation are plotted in the {\em lower} panel. Dashed lines show $1\sigma$ deviations from the mean $M_{Y_{SZ}}$--$M_{Y_X}$ relation. The fact that clusters that are outliers in both relations 
(those with overlaying circle and triangle) 
do not have consistent mass estimates within $1\sigma$ indicates that the errors in $M_{Y_{SZ}}$ and $M_{Y_X}$ are not correlated.}
\label{my_myx}
\end{center}
\end{figure}

Figure~\ref{my_myx} shows the mass predicted by the $Y_{SZ}$--$M$ relation versus that predicted by the $Y_X$--$M$ relation ($M$ is the true mass). Clusters in the upper panel have less than $2\sigma$ deviations from both the mean $Y_{SZ}$--$M$ relation and $Y_X$--$M$ relation. The lower panel shows the clusters whose mass scatter is bigger than $2\sigma$ for either relation. From the upper panel we can see that clusters that have consistent $M_{Y_{SZ}}$ and $M_{Y_X}$ are mostly faithful tracers of their true masses. But how about the outliers in both the true $Y_{SZ}$--$M$ and $Y_X$--$M$ relations? If they give consistent mass estimates, then there would be a similar problem of correlated error as described above. Fortunately, we find that the outliers in both relations (those plotted with both open and filled symbols) would {\it not} yield consistent mass estimates. This is because while the $Y_{SZ}$--$M$ outliers are due to cluster morphology, as discussed in \S~\ref{subsec:morphology}, we find that the $Y_X$--$M$ outliers are primarily dynamically unrelaxed clusters. Since the errors come from different physical sources, they are not correlated. 

This implies the possibility of cutting off the outliers by selecting only clusters whose $M_{Y_{SZ}}$ and $M_{Y_X}$ agree within $1\sigma$. Moreover, applying the same cut will also remove almost all the other $Y_{SZ}$--$M$ outliers. That is, among the 21 $Y_{SZ}$--$M$ outliers (15 are $2\sigma$ and 6 are $3\sigma$), 19 of them (13 are $2\sigma$ and 6 are $3\sigma$) can be ruled out using this method. After applying the cut, we find that the RMS scatter in $Y_{SZ}$--$M$ is reduced from $12.07\%$ to $8.77\%$ (i.e.\ by $27.3\%$), and also the non-lognormality of the $Y_{SZ}$--$M$ scatter is greatly reduced (skewness reduced from 0.82 to 0.30; kurtosis from 2.17 to -0.07). Therefore, combining mass estimates from $Y_X$ measurements may be an effective way of both reducing the scatter and removing $Y_{SZ}$--$M$ outliers.
Note that since the projection effect is not included in our simulation, we expect there would be more $Y_{SZ}$--$M$ outliers in reality, while the $Y_X$--$M$ relation is relatively insensitive to the projections. Indeed, the contamination by projection errors estimated by \cite{santafe06} using a light cone simulation is $\sim 25\%$ (for a projected radius of $R_{500}$), larger than ours (21 out of 619 clusters). However, because of the fact that the errors in the $Y_{SZ}$--$M$ and $Y_X$--$M$ relations are not correlated, clusters that are subject to projection errors can also be removed using the same method.



\section{Discussion and Conclusions}
\label{Sec:conclusion}

Galaxy clusters are invaluable cosmological probes. Accurate measurement of cluster masses is crucial and often relies on the mass-observable relations. However, to constrain the cosmological parameters at the few percent level, the systematics and scatter in these relations must be thoroughly understood. In this work we investigated the sources of intrinsic scatter in the SZ flux-mass ($Y$--$M$) relation using a hydrodynamics plus $N$-body simulation of galaxy clusters within a cosmological volume. 
Exploring the origin of the intrinsic scatter not only provides physical insights into the formation of galaxy clusters, but also has two main advantages for using clusters in cosmology. The first is that it allows us to avoid possible systematic biases in the derived cosmological constraints. Do mergers bias the scaling relation? Is the intrinsic scatter lognormal? What are the gains and issues of combining SZ and X-ray scaling relations? Secondly, if we understand the sources of scatter, it is possible to reduce the scatter by removing the contribution from a certain source (see \S~\ref{subsec:cc} for details), and thus tighten the scaling relation to obtain better estimates of cluster masses.

To address these questions, we derived the scatter around the best-fit $Y$--$M$ relation from the simulated clusters. We first assessed the lognormality of scatter by computing the skewness ($\gamma$) and kurtosis ($\kappa$) of the scatter distribution. Then we investigated the possible sources of scatter, including halo concentrations, dynamical states and cluster morphology, by correlating the scatter with quantitative measures of each source. We also constructed a set of idealized cluster samples with varied assumptions about the sources of scatter to decompose the percentage contribution from each effect. Finally we compared cluster masses derived from the SZ flux and from the low-scatter X-ray mass proxy, $Y_X$, and examined whether such consistency checks can help rule out outliers in the true $Y_{SZ}$--$M$ and $Y_X$--$M$ relations, or whether issues like correlated errors would affect the accuracy when combining SZ and X-ray scaling relations. Our main results are summarized below.

1.\ The RMS scatter in the $Y$--$M$ relation is $\sim 5-15\%$ and decreases with cluster masses and redshifts. We find that the scatter in our simulation can be expressed in the functional form, $\sigma(M,z) = A \log M + B \log(1+z) + C$ (Eq.\ \ref{eq:scatter_fit}), where the redshift evolution is equivalent to re-scaling with respect to the characteristic mass scale in the self-similar model.

2.\ The distribution of the $Y$--$M$ scatter is non-lognormal with positive skewness and kurtosis across a wide range of different limiting masses and redshifts, because of the limited number of clusters at the higher-mass end and the tail in the scatter distribution due to morphology at the lower-mass end. 

3.\ There is a strong correlation between the $Y$--$M$ scatter and the concentration, which can be used to reduce the $Y$--$M$ scatter from 12.07\% to 7.34\% (i.e.\ by 38.9\%). 

4.\ The correlation between the scatter and cluster dynamical state is weak. Though merger boosts and departure from hydrostatic equilibrium can partly drive the dispersion, the net effect is that mergers do not cause a significant bias in the scaling relation.

5.\ There is a moderate trend that clusters that are more elongated along the line of sight tend to scatter high. More importantly, they are the main outliers that cause the non-lognormality of scatter.

6.\ By decomposing the scatter using the idealized cluster samples, we find the percentage contribution from each source of scatter: $\sim 10\%$ due to variations in morphology, $\sim 10-20\%$ due to variations in concentration (under the assumption of spherical symmetry and hydrostatic equilibrium), $\sim 10-15\%$ due to departure from hydrostatic equilibrium, $\sim 30-60\%$ due to merger boosts, $\sim 0-10\%$ from variations in gas fractions. The remainder (due to unaccounted-for sources) is $\sim 0-30\%$.     

7.\ We find that the RMS scatter in $Y_{SZ}$--$M$ is reduced from $12.07\%$ to $8.77\%$ (i.e.\ by $27.3\%$) when X-ray measurements are combined with SZ.

8.\ The errors in mass determined using $Y_{SZ}$ and $Y_X$ come from different causes. Therefore, excluding clusters with inconsistent estimates can effectively remove the outliers in both $Y_{SZ}$--$M$ and $Y_X$--$M$ relations, especially $Y_{SZ}$--$M$ outliers that are subject to projection errors.

In our current simulation, radiative cooling and heating mechanisms are not included, since we would like to disentangle the scatter driven by the gravitational effects from other baryonic physics that are not fully understood. Moreover, it has been shown that the integrated SZ flux, and more specifically the scatter, slope, and redshift evolution of the $Y$--$M$ relation, are generally insensitive to details of cluster gas physics \citep{2004MNRAS.348.1401D, 2005ApJ...623L..63M, 2006ApJ...650..538N}. 
Since the non-lognormality is mainly caused by the effects of projection and cluster morphology, we could assess the potential impact of baryonic physics on these two sources. \cite{shaw07} showed that the influence of different gas physics on the properties of large-scale projections is negligible. Recently \cite{2010arXiv1003.2270L} has reported the difference in cluster shapes between simulations with and without cooling and star formation. They found that in the cooling plus star formation simulation clusters are more spherical outside the core ($r > 0.1R_{500}$) but more triaxial inside the core. It is difficult to estimate directly from their results how much this difference in morphology would affect the non-lognormality of the $Y$--$M$ scatter. However, as pointed out by \cite{2010arXiv1003.2270L}, their simulation may suffer from the overcooling problem and hence their results can be considered as an upper limit. Moreover, \cite{2006ApJ...650..538N} used clusters from the same simulations and showed that when the SZ flux is integrated to $R_{500}$, the $Y$--$M$ scatter is insensitive to the gas physics included. Therefore, we expect the effect of gas physics on the non-lognormality estimated in this paper, if any, is very small. We will present a more detailed comparison in a separate paper.

Our results have several important implications for cluster cosmology. First of all, the strong correlation with halo concentrations can be used for observed clusters to reduce the scatter in the scaling relations for better mass estimates. Potentially this method can be applied to any observable for which such a correlation exists, such as gas fractions (which is expected to play a more important role when including other baryonic physics, see \cite{2010ApJ...715.1508S}). Secondly, the weak influence of mergers is good news for cluster cosmology, because it implies that when deriving observed scaling relations, it is unnecessary to worry much about the selection bias due to the impact of mergers. Finally, the non-lognormality of the $Y$--$M$ scatter has an impact on cosmological constraint studies. As demonstrated by \cite{shaw10}, both positive skewness and kurtosis cause up-scattering of clusters and thus would increase cluster counts above a given limiting mass, which is equivalent to an increase in the amount of scatter. For SZ surveys like the South Pole Telescope (SPT) survey, the skewness and kurtosis of the intrinsic scatter have to be less than 0.5 to ensure that uncertainty in the amount of scatter does not degrade the constraint on the dark energy equation of state $w$. However, we find that the intrinsic skewness and kurtosis can be much greater than 0.5 across a wide range of limiting masses and redshifts. These values are very likely to be lower limits because the projection effect of large-scale structure is absent in our analysis. Therefore, our results suggest that the assumption of lognormal scatter is inappropriate for scaling relations like the $Y$--$M$ relation whose scatter is easily skewed by cluster morphology, projection effects, etc. Instead, in self-calibration studies of cosmological constraints that require assumptions about the form of scatter, it is necessary to include the higher-order moments in the parametrization. During the next decade, more and more data from multi-wavelength cluster observations will be available. We expect that more detailed studies of the intrinsic scatter in the scaling relations will continue to yield essential information both for cluster physics and cluster cosmology.


\acknowledgments

The authors acknowledge support under a Presidential Early Career Award from the U.S.
Department of Energy, Lawrence Livermore National Laboratory (contract B532720).
Additional support was provided by NASA Headquarters under the NASA Earth and
Space Science Fellowship Program (NNX08AZ02H). 
SB acknowledges support from the LDRD and IGPP program at Los Alamos National Laboratory.
The work described here was carried out using the resources of the National Center for Supercomputing Applications (allocation MCA05S029)
and the National Center for Computational Sciences at Oak Ridge National Laboratory (allocation AST010). FLASH was developed largely by the DOE-supported ASC/Alliances Center for Astrophysical Thermonuclear Flashes at the University of Chicago.


\bibliography{scatter}

\end{document}